\documentclass[10pt,conference]{IEEEtran}

\IEEEoverridecommandlockouts\IEEEpubid{\makebox[\columnwidth]{ 978-1-6654-3540-6/22~\copyright~2022 IEEE \hfill} \hspace{\columnsep}\makebox[\columnwidth]{ }}

\usepackage{amsmath,amssymb,amsfonts}
\usepackage{cite}
\usepackage{graphicx}
\usepackage{epstopdf}

\usepackage{booktabs}
\usepackage{graphicx}
\usepackage{textcomp}
\usepackage{xcolor}
\usepackage{times}
\usepackage{subfigure}         
\usepackage{amssymb,amsmath}
\usepackage{acronym}  
\usepackage{balance}

\usepackage{bm}         
\usepackage{algorithm} 
\usepackage{algorithmic}

\usepackage{lipsum}
\usepackage{stfloats}

\usepackage{setspace}

\usepackage{multirow}

\usepackage{url}

\usepackage{mathrsfs}

\usepackage{amsmath}
\allowdisplaybreaks[4]

\newtheorem{remark}{\bf Remark}

\newtheorem{lemma}{\bf Lemma}

\acrodef{lis}[LIS]{large intelligent surface}

\acrodef{ofdm}[OFDM]{orthogonal frequency division multiplexing}%
\acrodef{miso-ofdm}[MISO-OFDM]{multi-input single-output orthogonal frequency division multiplexing}%

\acrodef{ris}[RIS]{reconfigurable intelligent surface}%

\acrodef{irs}[IRS]{intelligent reflecting surface}%

\acrodef{qos}[QoS]{quality of service}%
\acrodef{idft}[IDFT]{inverse discrete Fourier transform}%
\acrodef{dft}[DFT]{discrete Fourier transform}%
\acrodef{cp}[CP]{cyclic prefix}%
\acrodef{csi}[CSI]{channel state information}%
\acrodef{awgn}[AWGN]{additive white Gaussian noise}%

\acrodef{qcqp}[QCQP]{quadratically constrained quadratic program}%
\acrodef{qp}[QP]{quadratic program}%
\acrodef{bs}[BS]{base station}%
\acrodef{qos}[QoS]{quality of service}%
\acrodef{ue}[UE]{user equipment}%
\acrodef{snr}[SNR]{signal-to-noise ratio}%
\acrodef{mmwave}[mmWave]{millimeter-wave}%

\acrodef{rf}[RF]{radio frequency}%

\acrodef{sinr}[SINR]{signal-to-interference-plus-noise ratio}%

\acrodef{ser}[SER]{symbol error rate}%

\acrodef{ace}[ACE]{adaptive cross-entropy}%
\acrodef{wsr}[WSR]{weighted sum-rate}%
\acrodef{udn}[UDN]{ultra-dense network}%

\def\BibTeX{{\rm B\kern-.05em{\sc i\kern-.025em b}\kern-.08em
		T\kern-.1667em\lower.7ex\hbox{E}\kern-.125emX}}

\begin{document}
	\title{Active RISs: Signal Modeling, Asymptotic Analysis, and Beamforming Design \vspace*{-0.4em}}
	\author{
		\IEEEauthorblockN{
			Zijian~Zhang\IEEEauthorrefmark{1},
			Linglong~Dai\IEEEauthorrefmark{1},~\IEEEmembership{Fellow,~IEEE}, {Xibi Chen}\IEEEauthorrefmark{1}, {Changhao Liu}\IEEEauthorrefmark{1}, {Fan Yang}\IEEEauthorrefmark{1},~\IEEEmembership{Fellow,~IEEE},\\{Robert Schober}\IEEEauthorrefmark{2},~\IEEEmembership{Fellow,~IEEE}, and {H. Vincent Poor}\IEEEauthorrefmark{4},~\IEEEmembership{Life Fellow,~IEEE}
		}
		\vspace{0.2cm}
		\IEEEauthorblockA{ 
			\IEEEauthorrefmark{1}Beijing National Research Center for Information Science and Technology (BNRist) \\
		 Department of Electronic Engineering,
			Tsinghua University, China\\	
			\IEEEauthorrefmark{2}Institute for Digital Communications, Friedrich-Alexander University Erlangen-Nürnberg, Germany\\
			\IEEEauthorrefmark{4}Department of  Electrical and Computer Engineering,  Princeton  University,  USA\\
			E-mails: zhangzj20@mails.tsinghua.edu.cn; daill@tsinghua.edu.cn; cxb17@tsinghua.org.cn; liuch17@tsinghua.org.cn;\\ fan\_yang@tsinghua.edu.cn; robert.schober@fau.de; poor@princeton.edu
			\vspace*{-1em}
		}
	}
	\maketitle

\begin{abstract}
	Reconfigurable intelligent surfaces (RISs) have emerged as a candidate technology for future 6G networks. However, due to the “multiplicative fading” effect, the existing passive RISs only achieve a negligible capacity gain in environments with strong direct links. In this paper, the concept of active RISs is studied to overcome this fundamental limitation. Unlike the existing passive RISs that reflect signals without amplification, active RISs can amplify the reflected signals via amplifiers integrated into their elements. To characterize the signal amplification and incorporate the noise introduced by the active components, we verify the signal model of active RISs through the experimental measurements on a fabricated active RIS element.
	Based on the verified signal model, we formulate the sum-rate maximization problem for an active RIS aided multi-user multiple-input single-output (MU-MISO) system and a joint transmit precoding and reflect beamforming algorithm is proposed to solve this problem. Simulation results show that, in a typical wireless system, the existing passive RISs can realize only a negligible sum-rate gain of 3\%, while the active RISs can achieve a significant sum-rate gain of 62\%, thus overcoming the “multiplicative fading” effect. Finally, we develop a 64-element active RIS aided wireless communication prototype, and the significant gain of active RISs is validated by field test.
\end{abstract}

\section{Introduction}	
From the first generation (1G) to 5G wireless communications, the wireless channel has been considered to be uncontrollable. Recently, due to the advances in meta-materials, \acp{ris} have been proposed \cite{Huang'18'2} for the purpose of intelligently controlling wireless channels to achieve improved communication performance. Specifically, an RIS is an array composed of a very large number of passive elements that reflects electromagnetic signals in a desired manner so as to reconfigure the propagation properties of wireless environment. As an important advantage of RIS, the negligible noise introduced by passive RISs enables a high array gain. Benefiting from this advantage, \acp{ris} are expected to introduce significant capacity gains in wireless systems \cite{Pan'19}. 

However, in practice, the expected capacity gains are typically only observed in communication environments where the direct link between transmitter and receiver is completely blocked or very weak. By contrast, in many scenarios where the direct link is not weak, conventional \acp{ris} can only achieve negligible capacity gains \cite{Najafi'20}. The reason behind this phenomenon is the “multiplicative fading” effect introduced by RISs, i.e., the equivalent path loss of the transmitter-RIS-receiver link is the product (instead of the sum) of the path losses of the transmitter-RIS and RIS-receiver links, which is usually thousands of times larger than that of the direct link \cite{Huang'18'2,Pan'19,Najafi'20}. As a result, the “multiplicative fading” effect makes it almost impossible for passive RISs to achieve noticeable capacity gains in many wireless environments. Therefore, to advance the practicability of RISs in future 6G wireless networks, a critical issue for RISs to be addressed is: \textit{How to break the fundamental performance bottleneck caused by the “multiplicative fading” effect?}
\par
To overcome the fundamental physical limitation of conventional passive RISs imposed by the “multiplicative fading” effect, in this paper, we investigate the concept of \textit{active} \acp{ris} to overcome the “multiplicative fading” effect. Different from the existing \textit{passive} \acp{ris} that passively reflect signals without amplification, the key feature of active \acp{ris} is their ability to actively reflect signals with amplification at the expense of additional power consumption. Firstly, through the experimental measurements on a fabricated active RIS element, we verify the signal model of active RISs, which characterizes the amplification of the incident signal and
accounts for the non-negligible thermal noise introduced
by the active elements. Based on the verified signal model, we further analyze the asymptotic performance of active RISs and formulate a sum-rate maximization problem for an active RIS aided multi-user multiple-input single-output (MU-MISO) system. Then, a joint transmit precoding and reflect beamforming algorithm is proposed to solve this problem. Simulation results show that, in a typical wireless system, the existing passive RISs achieve only a negligible sum-rate gain of 3\%, while the active RISs are able to achieve a substantial sum-rate gain of 62\%. Finally, we develop a 64-element active RIS aided wireless communication prototype, and field tests are conducted to validate the significant gain of active RISs.
\par
The rest of this paper is organized as follows.
In Section \ref{sec:model}, the concept of \acp{ris} and their signal models are introduced. In Section \ref{sec:PA}, the asymptotic performance of active RISs is analyzed. In Section \ref{sec:Precoding}, a sum-rate maximization problem is formulated, and a joint precoding and beamforming design is proposed to solve the problem. In Section \ref{sec:sim}, simulation results and experimental measurements are presented to validate the signal model and evaluate the performance of active RISs. Finally, conclusions are drawn in Section \ref{sec:con}. 

\section{Passive RISs and Active RISs}\label{sec:model}
\subsection{Conventional Passive RISs}\label{subsec:model1}
The RISs widely studied in most existing works are passive \cite{Huang'18'2,Pan'19,Najafi'20}. In general, each passive RIS element consists of a reflective patch terminated with an impedance-adjustable circuit for phase shifting. Thanks to the passive mode of operation, the thermal noise at passive RISs is usually negligible \cite{Pan'19}. Thereby, the signal model of an $N$-element passive RIS widely used in the literature is given as follows:
\begin{equation}\label{eqn:passive_model}
	\begin{aligned}
		{\bf{y}} = {\bf \Theta}{\bf{x}},
	\end{aligned}
\end{equation}
where ${\bf x}\in{\mathbb C}^N$ denotes the incident signal, ${\bf y}\in{\mathbb C}^N$ denotes the signal reflected by the RIS, and ${\bf \Theta}:={\rm diag}\left(e^{j\theta_1},\cdots,e^{j\theta_N}\right)\in{\mathbb C}^{N\times N}$ denotes the phase shift matrix of the RIS with ${\rm diag}(\cdot)$ being the diagonalization operation.
By properly adjusting ${\bf \Theta}$ to manipulate the $N$ signals 
reflected by the $N$ RIS elements to coherently add with the same phase at the receiver, a high array gain proportional to $N^2$ can be achieved, which is expected to significantly increase the \ac{snr} \cite{Huang'18'2} at the receiver.
\par
Unfortunately, this expected high capacity gain often cannot be realized in practice, especially in communication scenarios where the direct link between the transmitter and the receiver is strong. The reason for this negative result is the “multiplicative fading” effect introduced by passive RISs. Specifically, the equivalent path loss of the transmitter-RIS-receiver reflected link is the product (instead of the sum) of the path losses of the transmitter-RIS and RIS-receiver links, and therefore, it is thousands of times larger than that of the unobstructed direct link. Thereby, for an RIS to realize a noticeable capacity gain, thousands (or even millions) of RIS elements are required to compensate for this extremely large path loss \cite{Najafi'20}. The resulting high signaling overhead for channel estimation and the high complexity of real-time beamforming make the application of such a large number of passive RIS elements in practical wireless networks very challenging.

\subsection{Concept of Active RISs}\label{subsec:model2}

\begin{figure}[!t]
	\centering
	\includegraphics[width = 0.45\textwidth]{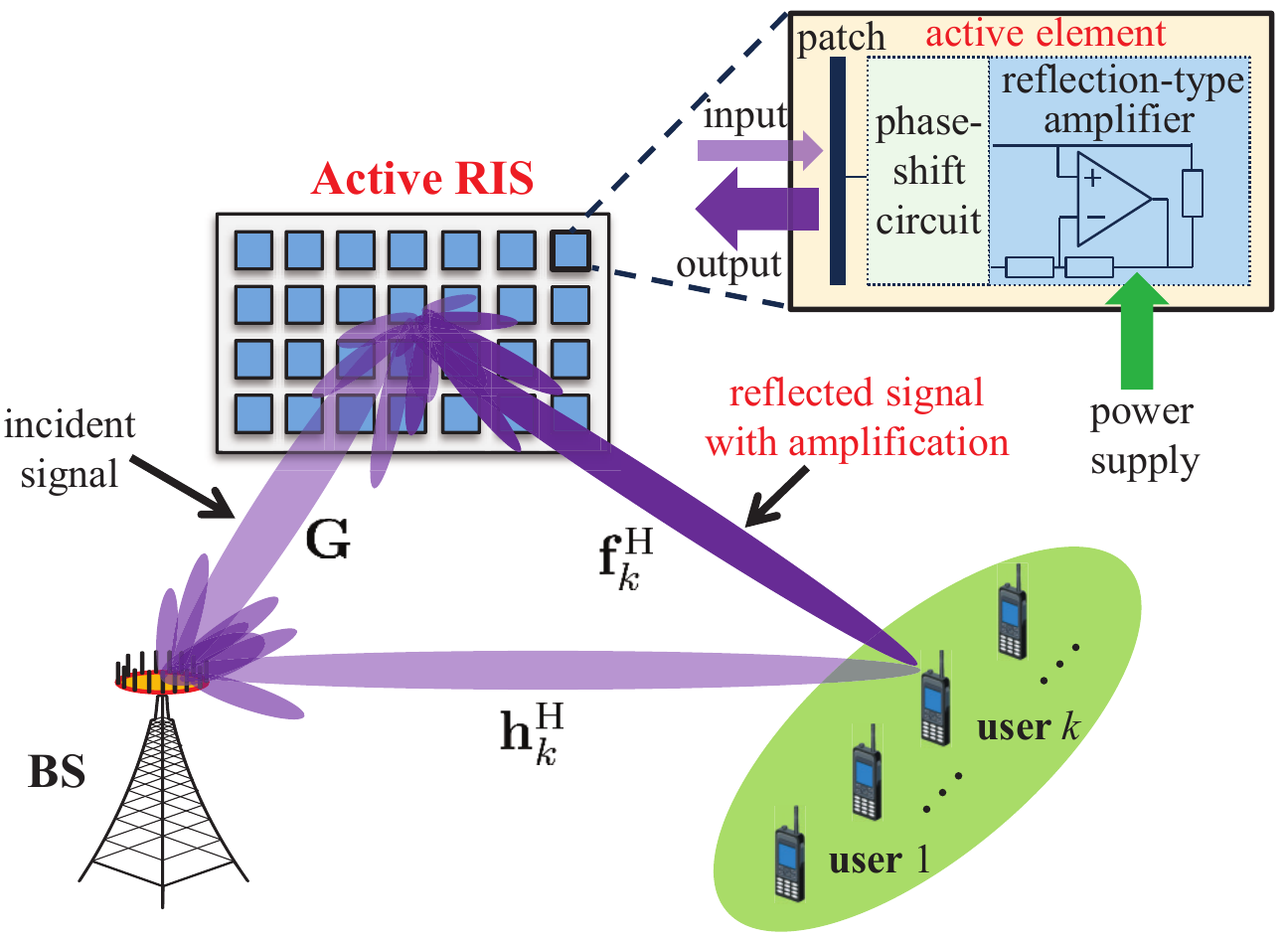}
	\vspace*{-0.5em}
	\caption{The downlink transmission in an active RIS aided MU-MISO system.
	}
	\label{img:model}
	\vspace*{-1em}
\end{figure}

To overcome the fundamental performance bottleneck caused by the “multiplicative fading” effect of RISs, we study the concept of active RISs as a promising solution\footnote{Note that active \acp{ris} are fundamentally different from relay-type \acp{ris} equipped with RF components and relays. Due to space constraints, we refer to the journal version of this paper for a more detailed discussion \cite[Remark 1]{Zhang'21}. 
}. As shown in Fig. \ref{img:model}, similar to the existing passive RISs, active RISs can also reflect the incident signals with reconfigurable phase shifts. Different from passive \acp{ris} that just reflect the incident signals without amplification, active \acp{ris} can further amplify the reflected signals. To achieve this goal, the key component of an active RIS element is the additionally integrated active reflection-type amplifier, which can be realized by different existing active components, such current-inverting converters or some integrated circuits \cite{Bousquet'12}. 

With reflection-type amplifiers supported by a power supply, the reflected and amplified signal of an $N$-element active RIS can be modeled as follows:
\begin{equation}\label{eqn:active_model}
	\begin{aligned}
		{\bf{y}} = \underbrace{{\bf{P}}{\bf \Theta}{\bf{x}}}_{\text{Desired signal}} + \underbrace {{\bf{P}}{\bf \Theta} {\bf{v}}}_{\text{Dynamic noise}} +\underbrace {{{\bf{n}}_{\text{s}}}}_{\text{Static noise}},
	\end{aligned}
\end{equation}
where ${\bf{P}}:={\rm diag}\left(p_1,\cdots,p_N\right)\in{\mathbb R}^{N\times N}$ denotes the amplification factor matrix of the active RIS, wherein each element $p_n$ can be larger than one thanks to the integrated reflection-type amplifier. Due to the use of active components, active RISs consume additional power for amplifying the reflected signals, and the thermal noise introduced by active RIS elements cannot be neglected as is done for passive RISs. Particularly, as shown in (\ref{eqn:active_model}), the introduced noise can be classified into dynamic noise and static noise \cite{Bousquet'12}. Specifically, ${\bf{v}}$ is related to the input noise and the inherent device noise of the active RIS elements \cite{Bousquet'12}, while the static noise ${\bf{n}}_{\text{s}}$ is unrelated to ${\bf{P}}$ and is usually negligible compared to the dynamic noise ${{\bf{P}}{\bf \Theta} {\bf{v}}}$, as will be verified by experimental results in Section \ref{sub:vr:sm}. Thus, here we neglect  ${\bf{n}}_{\text{s}}$ and model ${\bf{v}}$ as ${\bf{v}} \sim \mathcal{C} \mathcal{N}\left(\mathbf{0}_N, \sigma _{v}^2{\bf{I}}_N\right)$, where $\mathcal{C} \mathcal{N}\!\left({\bm \mu}, {\bf \Sigma } \right)$ denotes the complex multivariate Gaussian distribution with mean ${\bm \mu}$ and variance ${\bf \Sigma }$, $\mathbf{I}_{L}$ is an $L\times L$ identity matrix, and $\mathbf{0}_{L}$ is an $L\times 1$ zero vector.

\subsection{Active RIS Aided  MU-MISO System}\label{subsec:model3}
Consider an active RIS aided downlink MU-MISO system as shown in Fig. \ref{img:model}, where an $M$-antenna \ac{bs} serves $K$ single-antenna users simultaneously with the aid of an $N$-element active \ac{ris}. Let $\mathbf{s} :=\left[s_{1}, \cdots, s_{K}\right]^{\rm T} \in \mathbb{C}^{K}$ denote the transmitted symbol vector for the $K$ users and let ${\bf{w}}_k\in{\mathbb C}^{M\times 1}$ denote the \ac{bs} precoding vector for symbol $s_k$. According to  (\ref{eqn:active_model}), signal $r_k\in\mathbb{C}$ received at user $k$ can be modeled as follows:
\begin{align}\label{eqn:signal}
	{r_k}
	&= (\underbrace {{\bf{h}}_k^{\rm H}}_{{\text{Direct link}}} \!\! + \underbrace {{\bf{f}}_k^{\rm H}{\bf{P}}{{\bf{\Theta }}}{\bf{G}}}_{{\text{Reflected link}}})\sum\nolimits_{j = 1}^K \! {{{\bf{w}}_j}{s_j}} +\!\!\!\!\!\!\!\! \underbrace{{\bf{f}}_k^{\rm H}{{\bf{P}}}{{\bf{\Theta }}}{\bf{v}}}_{\text{Noise introduced by active RIS}}  \notag \\ &~~~~~~~~+\underbrace {z_k}_{\text{Noise introduced at user $k$}},
\end{align}
where ${[\cdot]^{\rm H}}$ denotes the conjugate-transpose operation; ${\bf{G}}\in\mathbb{C}^{N\times M}$, ${\bf{h}}_k^{\rm H}\in\mathbb{C}^{1\times M}$, and ${\bf{f}}_k^{\rm H}\in\mathbb{C}^{1\times N}$ characterize the channels between the \ac{bs} and the RIS, between the \ac{bs} and user $k$, and between the \ac{ris} and user $k$, respectively; and $z_k$ denotes the \ac{awgn} at user $k$ with zero mean and variance $\sigma^2$.


\section{Performance Analysis}\label{sec:PA}
In this section, we analyze the performance of active RISs to reveal their notable capacity gains compared to passive RISs. To this end, in order to make the problem analytically tractable and get insightful results, in this section, we consider a single-user single-input single-output (SU-SISO) system with $M=1$ \ac{bs} antenna and $K=1$ user, while the general MU-MISO case is studied in Section \ref{sec:Precoding}.
\subsection{Asymptotic \ac{snr} for Passive RISs and Active RISs}\label{subsec:PA1}
To illustrate the capacity gain provided by passive/active RIS aided reflected links, for the moment, we ignore the direct link by setting ${\bf h}_k$ to zero, as was done in, e.g., \cite{Wu'19}. Furthermore, for simplicity, we assume that each active RIS element has the same amplification factor (i.e., $p_n:= p$). For a fair comparison with the asymptotic performance of passive RISs, similar to \cite{Wu'19}, we assume Rayleigh-fading channels.

We first redefine the \ac{bs}-RIS channel matrix and the RIS-user channel vector as ${\bf G}:={\bf g}\in{\mathbb C}^{N\times 1}$ and ${\bf f}_k:={\bf f}\in{\mathbb C}^{N\times 1}$, respectively. Then, we recall the following lemma from \cite{Wu'19} for the asymptotic \ac{snr} achieved by passive RISs.
\begin{lemma}[Asymptotic \ac{snr} for passive RISs]
	Assuming ${\bf{f}} \sim \mathcal{C} \mathcal{N}\left(\mathbf{0}_N, \varrho_{f}^{2} {\bf{I}}_N\right)$, ${\bf{g}} \sim \mathcal{C} \mathcal{N}\left(\mathbf{0}_N, \varrho_{g}^{2} {\bf{I}}_N\right)$ and letting $N \to \infty$, the asymptotic \ac{snr} $\gamma_{\text{passive}}$ of a passive RIS aided SU-SISO system is given by
	\begin{align}\label{eqn:ag_passive}
		\gamma_{\text{passive}}  \to {N^2}\frac{{P_{{\text{BS}}}^{{\max}}{\pi ^2}\varrho _f^2\varrho _g^2}}{{16{\sigma ^2}}},
	\end{align}
	where ${P^{\max}_{\text{BS}}}$ denotes the maximum transmit power at the \ac{bs}.
\end{lemma}
\begin{IEEEproof}
	The proof can be found in \cite[Proposition 2]{Wu'19}.
\end{IEEEproof}

For comparison, under the same transmission conditions, we provide the asymptotic \ac{snr} of an active RIS aided SU-SISO system in the following lemma.

\begin{lemma}[Asymptotic \ac{snr} for active RISs]
	Assuming ${\bf{f}} \sim \mathcal{C} \mathcal{N}\left(\mathbf{0}_N, \varrho_{f}^{2} {\bf{I}}_N\right)$, ${\bf{g}} \sim \mathcal{C} \mathcal{N}\left(\mathbf{0}_N, \varrho_{g}^{2} {\bf{I}}_N\right)$ and letting $N \to \infty$, the asymptotic \ac{snr} $\gamma_{\text{active}}$ of an active RIS aided SU-SISO system is given by
	\begin{align}\label{eqn:ag_active}
		\gamma_{\text{active}}  \to N \frac{{P_{{\text{BS}}}^{{\max}}P_{\text{A}}^{{\max}}{\pi ^2}\varrho _f^2\varrho _g^2}}{{16\left( {P_{\text{A}}^{{\max}}\sigma _v^2\varrho _f^2 + P_{{\text{BS}}}^{{\max}}{\sigma ^2}\varrho _g^2 + {\sigma ^2}\sigma _v^2} \right)}},
	\end{align}
	where ${P^{\max}_{\text{A}}}$ denotes the maximum reflect power of the active \ac{ris}.
\end{lemma}
\begin{IEEEproof}
	Please see the journal version \cite[Appendix A]{Zhang'21}.
\end{IEEEproof}
\begin{remark}
	From (\ref{eqn:ag_active}) we observe that, the asymptotic \ac{snr} of an active RIS aided SU-SISO system depends on both the \ac{bs} transmit power $P_{{\text{BS}}}^{{\max}}$ and the reflect power of the active \ac{ris} $P_{\text{A}}^{{\max}}$. When  $P_{{\text{BS}}}^{{\max}} \to \infty$, the asymptotic \ac{snr} will be upper-bounded by ${\gamma _{{\text{active}}}} \to NP_{\rm{A}}^{\max }{\pi ^2}\varrho _f^2/\left( {16{\sigma ^2}} \right)$, which is independent of the \ac{bs}-RIS channel $\bf g$ and the noise power at the active RIS ${\sigma_v^2}$. Similarly, if $P_{\rm{A}}^{\max } \to \infty$, the asymptotic \ac{snr} will be upper-bounded by ${\gamma _{{\text{active}}}} \to NP_{{\rm{BS}}}^{\max }{\pi ^2}\varrho _g^2/16\sigma _v^2$, which is independent of the RIS-user channel $\bf f$ and the noise power at the user ${\sigma^2}$. These results reveal that, to increase the sum-rate of active RIS aided systems, the negative impact of small $\bf g$ and large ${\sigma _v^2}$ on system performance can be alleviated by increasing the \ac{bs} transmit power $P_{{\text{BS}}}^{{\max}}$, and the negative impact of small $\bf f$ and large ${\sigma^2}$ can be reduced by increasing the reflect power of the active RIS $P_{\text{A}}^{{\max}}$. 
\end{remark}
\subsection{Comparisons between Passive RISs and Active RISs}\label{subsec:PA2}

We can observe from {\it Lemma 1} and {\it Lemma 2} that, compared to the asymptotic \ac{snr} for passive RISs $\gamma_{\text{passive}}$ in (\ref{eqn:ag_passive}) which is proportional to $N^2$, the asymptotic \ac{snr} for active RISs $\gamma_{\text{active}}$ in (\ref{eqn:ag_active}) is proportional to $N$ due to the noises introduced by the use of active components. At first glance, it seems that the \ac{snr} achieved by passive RISs $\gamma_{\text{passive}}$ always exceeds the \ac{snr} achieved by active RISs $\gamma_{\text{active}}$. However, this is actually not the case. The reason behind this counterintuitive behavior is that, due to the large path loss caused by the ``multiplicative fading'' effect and thanks to the use of the reflection-type amplifiers in active RISs, only when $N$ is unaffordably large can passive RISs outperform active RISs.

To illustrate this claim, let us consider two different SU-SISO systems, which are aided by an active RIS and a passive RIS, respectively. Then, the following lemma specifies the condition that has to be met for passive RISs to outperform active RISs.
\begin{lemma}[Case when passive RISs outperform active RISs]
	Assuming the number of RIS elements $N$ is large, the required number of elements $N$ for a passive RIS to outperform an active RIS has to satisfy
	\begin{align}\label{eqn:N_>}
		N \ge \frac{{P_{{\text{BS-A}}}^{{\max}}}}{{P_{{\text{BS-P}}}^{{\max}}}}\frac{{P_{\text{A}}^{{\max}}{\sigma ^2}}}{{\left( {P_{\text{A}}^{{\max}}\sigma _v^2\varrho _f^2 + P_{{\text{BS-A}}}^{{\max}}{\sigma ^2}\varrho _g^2 + {\sigma ^2}\sigma _v^2} \right)}},
	\end{align}
	where ${P^{{\max}}_{\text{BS-A}}}$ denotes the maximum BS transmit power for the active RIS aided system and ${P^{{\max}}_{\text{BS-P}}}$ denotes that for the passive RIS aided system.
\end{lemma}
\begin{IEEEproof}
	Please see the journal version \cite[Appendix B]{Zhang'21}.
\end{IEEEproof}

Next, we consider a specific setup to compare the user's achievable \acp{snr} in the above two systems. For a fair comparison, we constrain the total power consumption $P^{\max}$ of the two systems to $2$ W by setting $P_{\text{BS-P}}^{\max}=2$ W for the passive RIS aided system and $P_{\text{BS-A}}^{\max}=P_{\text A}^{\max}=1$ W for the active RIS aided system, respectively. Therefore, when $\sigma^2=\sigma _{v}^2=-70$ dBm and ${\varrho _f^2}={\varrho _g^2}=-70$ dB, the required number of elements $N$ for the passive RIS to outperform the active RIS is $2.5\times 10^6$ according to (\ref{eqn:N_>}), which is impractical to realize with current technology. Conversely, for a more practical number of elements of $N=256$, according to (\ref{eqn:ag_active}) and (\ref{eqn:ag_passive}), the \ac{snr} achieved by the passive RIS is $\gamma_{\text{passive}}\approx9.0$ dB, while the \ac{snr} achieved by the active RIS is $\gamma_{\text{active}}\approx49.0$ dB, which is about $10,000$ times higher than $\gamma_{\text{passive}}$. 

\section{Joint Transmit Precoding and Reflect Beamforming Design}\label{sec:Precoding}
To investigate the capacity gain enabled by the use of active RISs in typical wireless communication scenarios, in this section, we consider more general MU-MISO systems.
According to the model in (\ref{eqn:signal}), the \ac{sinr} at user $k$ can be obtained as
\begin{equation}
	{\gamma _k} = \frac{{{{\left| {{{\bf{\bar h}}^{\rm H}_k}{{\bf{w}}_k}} \right|}^2}}}{{\sum\nolimits_{j = 1,j \ne k}^K {{{\left| {{{\bf{\bar h}}^{\rm H}_k}{{\bf{w}}_j}} \right|}^2}}  + {{\left\| {{{\bf{f}}_k^{\rm H}}{{\bf{P}}}{{\bf{\Theta }}}} \right\|}^2}\sigma _{v}^2 + {\sigma ^2}}},
\end{equation}
wherein ${{\bf{\bar h}}^{\rm H}_k} = {{\bf{h}}^{\rm H}_k} + {{\bf{f}}_k}^{\rm H}{{\bf{P}}}{{\bf{\Theta }}}{\bf{G}}\in{\mathbb{C}^{1\times M}}$
is the equivalent channel from the \ac{bs} to user $k$, which includes both the direct link and the reflected link. Therefore, the original problem of sum-rate maximization, subject to the power constraints at the \ac{bs} and the active RIS, can be formulated as follows:
\begin{subequations}\label{eqn:problem}
	\begin{align}
		\!\!\!\!{\cal P}_o: \mathop {\max }\limits_{{\bf{w}},{\bf{P}},{\bf{\Theta }}} &~ R_{\rm{sum}}({\bf{w}},{\bf{P}},{\bf{\Theta }})= \sum\nolimits_{k = 1}^K {{{\log }_2}\left( {1 + {\gamma _k}} \right)}, \label{eqn:sum-rate} \\
		{\rm s.t.}\,\, &{\rm C_1}\!:\sum\nolimits_{k = 1}^K {{{\left\| {{{\bf{w}}_k}} \right\|}^2}} \le {P^{\max}_{\text{BS}}},  \\
		&{\rm C_2}\!: \sum\nolimits_{k = 1}^K\! {{{\left\| {{{\bf{P}}}{{\bf{\Theta }}}{\bf{G}}{{\bf{w}}_k}} \right\|}^2}  \!\!+\! \left\| {{{\bf{P}}}{{\bf{\Theta }}}} \right\|^2\sigma_{v}^2} \!\le\! {P^{\max}_{\text{A}}}, 	 
	\end{align}
\end{subequations}
where ${\bf{w}} := {\left[ {{\bf{w}}_1^{\rm T}, \cdots ,{\bf{w}}_K^{\rm T}} \right]^{\rm T}}$ is the overall transmit precoding vector for the $K$ users; ${\rm C_1}$ and ${\rm C_2}$ are the power constraints at the BS and active RIS, respectively. Due to the non-convexity and the highly coupled variables in problem ${\cal P}_o$ in (\ref{eqn:problem}), the joint design of ${\bf{w}}$, ${\bf{P}}$, and ${\bf{\Theta}}$ is challenging.

To efficiently solve the above problem, we develop a joint precoding and beamforming algorithm based on alternating optimization and fractional programming (FP). Note that ${\bf{P}}$ and ${\bf \Theta}$ always appear in product form in problem ${\cal P}_o$ in (\ref{eqn:problem}). Therefore, ${\bf P}$ and ${\bf \Theta}$ can be merged as ${\bf{\Psi }}={\bf P}{\bf \Theta}={\rm diag}\left(p_1e^{j\theta_1},\cdots,p_Ne^{j\theta_N}\right)\in{\mathbb C}^{N\times N}$. We refer to $\bf{\Psi }$ as the RIS beamforming matrix. Next, to deal with the non-convex sum-of-logarithms and fractions in (\ref{eqn:problem}), we exploit the FP methods proposed in \cite{Shen'18'1} to decouple the variables in problem ${\cal P}_o$ in (\ref{eqn:problem}). This leads to the following lemma.
\begin{lemma}[Equivalent problem for sum-rate maximization]
	By introducing auxiliary variables ${\bm\rho}:=\left[\rho_1,\cdots,\rho_K\right]\in{\mathbb R}^{K}$ and  $\bm \varpi:=\left[\varpi_1,\cdots, \varpi_K\right]\in{\mathbb C}^{K}$, the original problem ${\cal P}_o$ in (\ref{eqn:problem}) can be equivalently reformulated as follows
		\begin{align}\label{eqn:problem_6}
			{\cal P}_1:	\mathop {\max }\limits_{{\bf{w}},{\bm{\Psi}}, {\bm{\rho}},{\bm\varpi}}  &~R_{\rm sum}'({\bf{w}},{\bm{\Psi}},{\bm \rho},{\bm \varpi}) = \sum\nolimits_{k = 1}^K {{{\ln }}\left( {1 + {\rho _k}} \right)}  - \notag \\ & ~~~~~~~\sum\nolimits_{k = 1}^K {\rho _k} +\sum\nolimits_{k = 1}^K {g({\bf{w}},{\bm{\Psi}},{\rho_k},{\varpi_k})}, \notag \\
			~~{\rm s.t.}~~ &{\rm C_1}, {\rm C_2},
		\end{align}
	where function $g({\bf{w}},{\bm{\Psi}},{\rho_k},{\varpi_k})$ is defined as
	\begin{equation}
		\begin{aligned}
			&g({\bf{w}},{\bm{\Psi}},{\rho_k},{\varpi_k})= 2\sqrt {\left( {1 + {\rho _k}} \right)} {\mathop{{\mathfrak R}}\nolimits} \left\{ {{\varpi_k ^*}{\bf{\bar h}}_k^{\rm H}{{\bf{w}}_k}} \right\} - \\ &~~~~~~~~~ {\left| \varpi_k \right|^2}\left( {\sum\nolimits_{j = 1}^K {{{\left| {{\bf{\bar h}}_k^{\rm H}{{\bf{w}}_j}} \right|}^2}}  + {{\left\| {{{\bf{f}}_k^{\rm H}}{{\bf{\Psi }}}} \right\|}^2}\sigma _{v}^2 + {\sigma ^2}} \right).
		\end{aligned}
	\end{equation}
\end{lemma}
\begin{IEEEproof}
	Constructive proof can be found in \cite[Subsection III-C]{Shen'18'1}.
\end{IEEEproof}
\par
Strong convergence of the FP methods was proved in \cite{Shen'18'1}. Thus, a locally optimal solution to (\ref{eqn:problem_6}) can be obtained by alternately optimizing the variables. For clarity, we summarize the proposed joint precoding and beamforming algorithm in {\bf Algorithm 1}, and the specific solutions for variables ${\bf{w}}$, ${\bm{\Psi}}$, ${\bm{\rho}}$, and ${\bm\varpi}$ are given in the following four steps, respectively.
\begin{algorithm}[!h] 
	\setstretch{1}
	\caption{Proposed joint transmit precoding and reflect beamforming algorithm} 
	\begin{algorithmic}[1] 
		\REQUIRE ~~ 
		Channels ${\bf{G}}$, ${\bf{h}}_k$, and ${\bf{f}}_k$, $\forall k\in\{1,\cdots,K\}$.
		\ENSURE ~~ 
		Optimized \ac{bs} precoding vector $\bf{w}$, amplification factor matrix of active RIS $\bf{P}$, phase shift matrix of active RIS $\bm{\Theta}$, and sum-rate $R_{\rm{sum}}$.	
		\STATE Randomly initialize $\bf{w}$, $\bf{P}$ and $\bf{\Theta}$;
		\WHILE {no convergence of $R_{\rm{sum}}$}
		\STATE Update ${\bm \rho}$ by (\ref{eqn:rho_update});
		\STATE Update $\bm{\varpi}$ by (\ref{eqn:update_varpi});
		\STATE Update ${\bf w}$ by solving problem ${\cal P}_2$ in (\ref{eqn:problem_9});		
		\STATE Update ${\bm \Psi}$ by solving problem ${\cal P}_3$ in (\ref{eqn:problem_10});
		\ENDWHILE	
		\STATE Obtain $\bf{P}$ and $\bm{\Theta}$ from ${\bm \Psi}$;
		\RETURN Optimized $\bf{w}$, $\bf{P}$, $\bm{\Theta}$, and $R_{\rm{sum}}$. 
	\end{algorithmic}
\end{algorithm}
\subsubsection{Fix $\left({\bf{w}},{\bf{\Psi}},{\bm \varpi}\right)$ and optimize ${\bm \rho}$} 
After fixing precoding vector ${\bf{w}}$, beamforming matrix ${\bf{\Psi}}$, and auxiliary variable ${\bm \varpi}$, the optimal $\bm \rho$ can be obtained by solving $\frac{{\partial {R_{{\rm{sum}}}'}}}{{\partial {\rho _k}}} = 0$ as 
\begin{equation}\label{eqn:rho_update}
	\begin{aligned}
		\rho^{\rm opt}_k  = \frac{{{\xi_k ^2} + \xi_k \sqrt {{\xi_k ^2} + 4} }}{2},\quad\forall k\in \{1,\cdots,K\},
	\end{aligned}
\end{equation}
where $\xi_k  = \Re \left\{ {\varpi _k^*{\bf{\bar h}}_k^{\rm H}{{\bf{w}}_k}} \right\}$.
\subsubsection{Fix $\left({\bf w},{\bf{\Psi}},{\bm \rho}\right)$ and optimize $\bm \varpi$}
After fixing the precoding vector ${\bf{w}}$, beamforming matrix ${\bf{\Psi}}$, and auxiliary variable ${\bm \rho}$, the optimal $\bm \varpi$ can be derived by solving $\frac{\partial {R_{{\rm{sum}}}'}}{{\partial \varpi_k }} = 0$ as
\begin{equation}\label{eqn:update_varpi}
	\begin{aligned}
		{\varpi_k^{{\rm{opt}}}} \!=&\!  \frac{{\sqrt {\left( {1 + {\rho _k}} \right)} {\bf{\bar h}}_k^{\rm H}{{\bf{w}}_k}}}{{\sum\nolimits_{j = 1}^K \!{{{\left| {{\bf{\bar h}}_k^{\rm H}{{\bf{w}}_j}} \right|}^2}}  \!+\! {{\left\| {{\bf{f}}_k^{\rm H}{{\bf{\Psi }}}} \right\|}^2}\sigma _{v}^2 + {\sigma ^2}}}.
	\end{aligned}
\end{equation}
\subsubsection{Fix $\left({\bf{\Psi}},{\bm \rho},{\bm \varpi}\right)$ and optimize ${\bf w}$}
To simplify the notations, we first introduce the following definitions:
\begin{subequations}\label{eqn:varibles}
	\begin{align}
		&{\bf{b}}_k^{\rm H} \!=\! \sqrt {\left( {1 + {\rho _k}} \right)} \varpi_k^*{\bf{\bar h}}_k^{\rm H},~
		{\bf b} = \left[{{\bf b}_{1}^{\rm T}},{{\bf b}_{2}^{\rm T}}, \cdots, {{\bf b}_{N}^{\rm T}}\right]^{\rm T},\\
		&\!\!\!{\bf{A}} \!=\! {{\bf{I}}_K} \!\otimes\! \sum\nolimits_{k = 1}^K \!\! {{\left| {{\varpi _k}} \right|^2}{{\bf{\bar h}}_k}{\bf{\bar h}}_k^{\rm H}},~
		{\bf{\Xi }} \!=\! {{\bf{I}}_K} \!\otimes \! \left( {{{\bf{G}}^{\rm H}}{{\bf{\Psi }}^{\rm H}}{\bf{\Psi G}}} \right),\\
		&\qquad \qquad \qquad P_{\text m}^{\max } = P_{{\text{A}}}^{{\max}} - {\left\| {{{\bf{\Psi }}}} \right\|^2}\sigma _{v}^2,
	\end{align}
\end{subequations}
where $\otimes$ denotes the Kronecker product. Then, problem ${\cal P}_1$ in (\ref{eqn:problem_6}) can be reformulated as follows:
\begin{equation}\label{eqn:problem_9}
	\begin{aligned}
		{\cal P}_2:~~	\mathop {\max }\limits_{{\bf{w}}} ~~ & {\mathop{\mathfrak R}\nolimits} \left\{2 {{{\bf{b}}^{\rm H}}{\bf{w}}} \right\} - {{\bf{w}}^{\rm H}}{\bf{Aw}},  \\
		{\rm s.t.}~~~ &{\rm C_1}: {{{\left\| {{{\bf{w}}}} \right\|}^2}} \le {P^{\max}_{\text{BS}}},\\
		\quad ~~~~ &{\rm C_2}: {{\bf{w}}^{\rm H}}{\bf{\Xi w}} \le P_{\text m}^{\max}.
	\end{aligned}
\end{equation}
Note that ${\cal P}_2$ in (\ref{eqn:problem_9}) is a standard quadratic constraint quadratic programming (QCQP) problem, which can be solved by alternating direction method of multipliers (ADMM).

\subsubsection{Fix $\left({\bf w},{\bm \rho},{\bm \varpi}\right)$ and optimize ${\bf{\Psi}}$}
Define ${\bm{\psi}}=\left[p_1e^{j\theta_1},\cdots,p_Ne^{j\theta_N}\right]^{\rm H}$ as the vectorized RIS beamforming matrix ${\bm{\Psi}}$, i.e., ${\rm diag}\left({\bm{\psi}}^{\rm H}\right):= {\bf{\Psi}}$.
While fixing ${\bf w}$ and ${\bm \rho}$ and ${\bm \varpi}$, problem ${\cal P}_1$ in (\ref{eqn:problem_6}) can be reformulated as follows:
\begin{equation}\label{eqn:problem_10}
	\begin{aligned}
		{\cal P}_3:~~\mathop {\max }\limits_{{\bm{\psi}}} \,\,\,\, &{\mathop{\mathfrak R}\nolimits} \left\{2 {{{\bm{\psi }}^{\rm H}}{\bm{\upsilon }}} \right\} - {{\bm{\psi }}^{\rm H}}{\bm{\Omega \psi }},  \\
		{\rm s.t.}~~ &{\rm C_2}: {{\bm{\psi }}^{\rm H}}{\bm{\Pi \psi }} \le P_{{\text{A}}}^{{\max}},
	\end{aligned}
\end{equation} 
wherein
\begin{subequations}
	\begin{align}
		&{{\bm{\upsilon }}} = \sum\nolimits_{k = 1}^K\sqrt {\left( {1 + {\rho _k}} \right)} {\rm{diag}}\left( {\varpi _k^*{\bf{f}}_k^{\rm H}} \right){\bf{G}}{{\bf{w}}_k} - \notag \\ &~~~~~~~\sum\nolimits_{k = 1}^K{\left| {{\varpi _k}} \right|^2}{\rm{diag}}\left( {{\bf{f}}_k^{\rm H}} \right){\bf{G}}\sum\nolimits_{j = 1}^K {{{\bf{w}}_j}{\bf{w}}_j^{\rm H}} {{\bf{h}}_k},\\
		&{{\bm{\Omega }}} = \sum\nolimits_{k = 1}^K{\left| {{\varpi _k}} \right|^2}{\rm{diag}}\left( {{\bf{f}}_k^{\rm H}} \right){\rm{diag}}\left( {{{\bf{f}}_k}} \right)\sigma _{v}^2 + \notag\\& \sum\nolimits_{k = 1}^K{\left| {{\varpi _k}} \right|^2}\sum\nolimits_{j = 1}^K\!\! {{\rm{diag}}\left( {{\bf{f}}_k^{\rm H}} \right){\bf{G}}{{\bf{w}}_j}{\bf{w}}_j^{\rm H}{{\bf{G}}^{\rm H}}{\rm{diag}}\left( {{{\bf{f}}_k}} \right)},\\
		&{\bf{\Pi }} = \sum\nolimits_{k = 1}^K {{\rm{diag}}\left( {{\bf{G}}{{\bf{w}}_k}} \right){{\left( {{\rm{diag}}\left( {{\bf{G}}{{\bf{w}}_k}} \right)} \right)}^{\rm H}}}  + \sigma _{v}^2{{\bf{I}}_N}.
	\end{align}
\end{subequations}

Note that problem ${\cal P}_3$ in (\ref{eqn:problem_10}) is also a standard QCQP problem. Thus, the optimal solution ${\bm{\psi}}^{\rm opt}$ can be obtained by adopting ADMM.

\section{Validation Results}\label{sec:sim}
\subsection{Validation Results for Signal Model}\label{sub:vr:sm}
\begin{figure*}[!t]
	\vspace*{3em}
	\centering
	\includegraphics[width=0.85\textwidth]{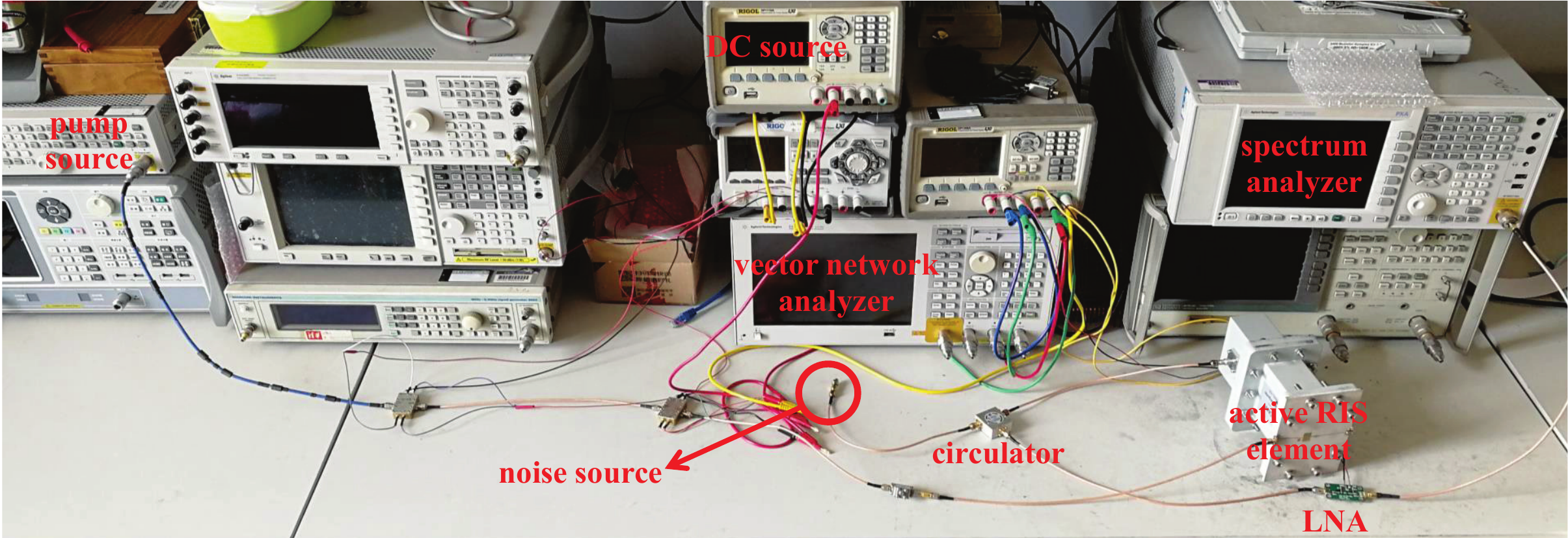}
	\vspace*{-1em}
	\caption{The experimental devices and environment used for validating the signal model (\ref{eqn:active_model}) of active RISs.}		
	\label{img:val:environment}
	\vspace*{-2em}
\end{figure*}
\begin{figure}[!t]
	\centering
	\includegraphics[width=0.45\textwidth]{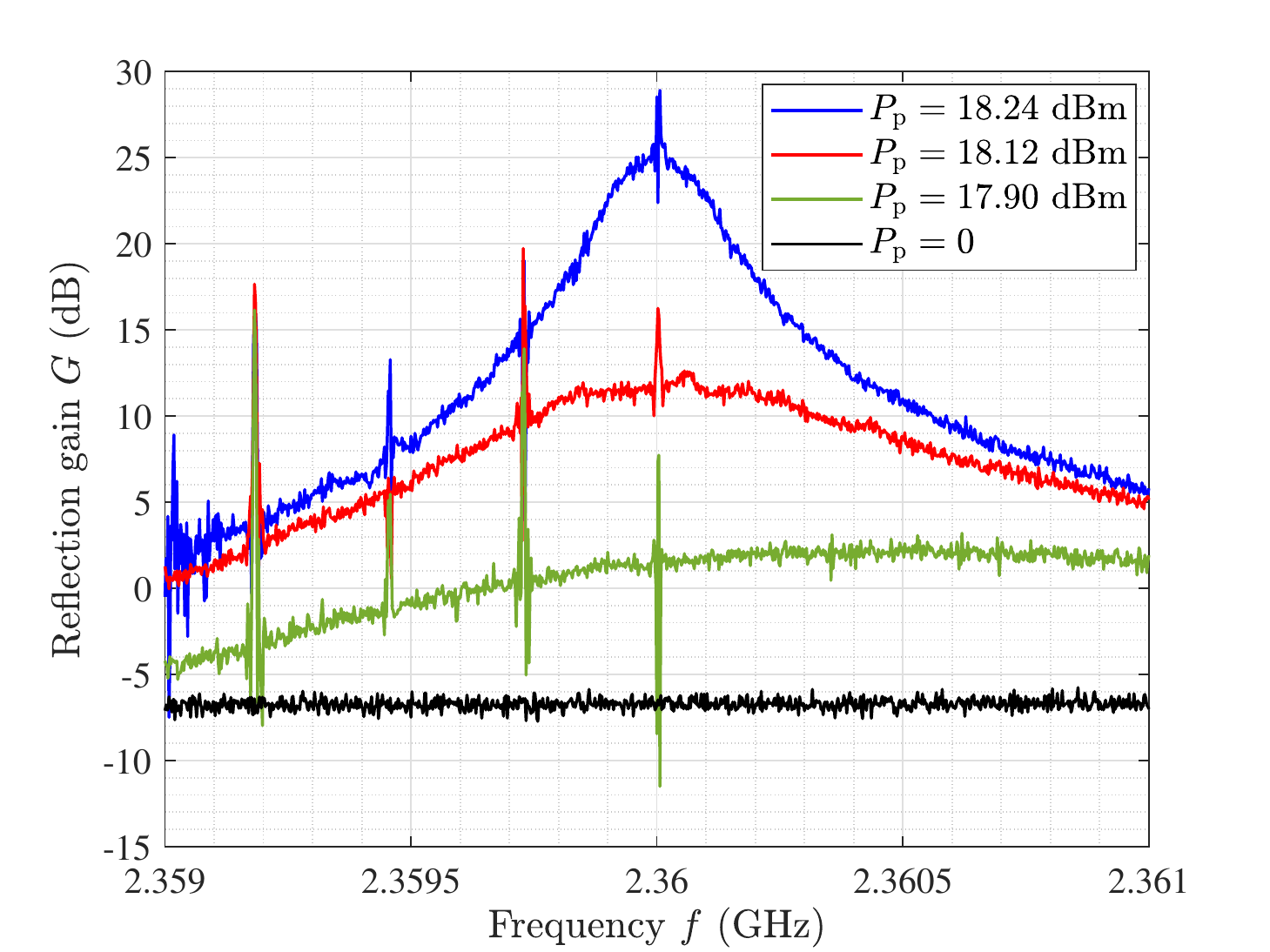}
	\vspace*{-1em}
	\caption{Experimental measurement result for reflection gain $G$ versus signal frequency $f$.}		
	\label{img:val:signal1}
	\vspace*{-1em}
\end{figure}
To validate the signal model (\ref{eqn:active_model}), we designed and fabricated an active RIS element with an integrated reflection-type amplifier for experimental measurements in \cite{Xibi'20}. Note that this design can be directly extended to the large-array case. Particularly, since the phase-shifting ability of \acp{ris} has been widely verified, we focus on studying the reflection gain and the noise introduced by an active RIS element. Thus, the validation of signal model (\ref{eqn:active_model}) is equivalent to validating
\begin{equation}\label{eqn:active_model_power}
	\begin{aligned}
		{P_y} = \underbrace {G{P_x}}_{{\text{Desired-signal power}}} + \underbrace {{G}\sigma _v^2 + \sigma _s^2}_{{\text{noise power}}},
	\end{aligned}
\end{equation}
where $P_y$ is the power of the signals reflected by the active RIS element; $P_x$ is the power of the incident signal; $ G:=p^2$ is the reflection gain of the active RIS element; ${G}\sigma _v^2$ and $\sigma _s^2$ are the powers of the dynamic noise and static noise introduced by the active RIS element, respectively.

\begin{figure}[!t]
	\centering
	\includegraphics[width=0.45\textwidth]{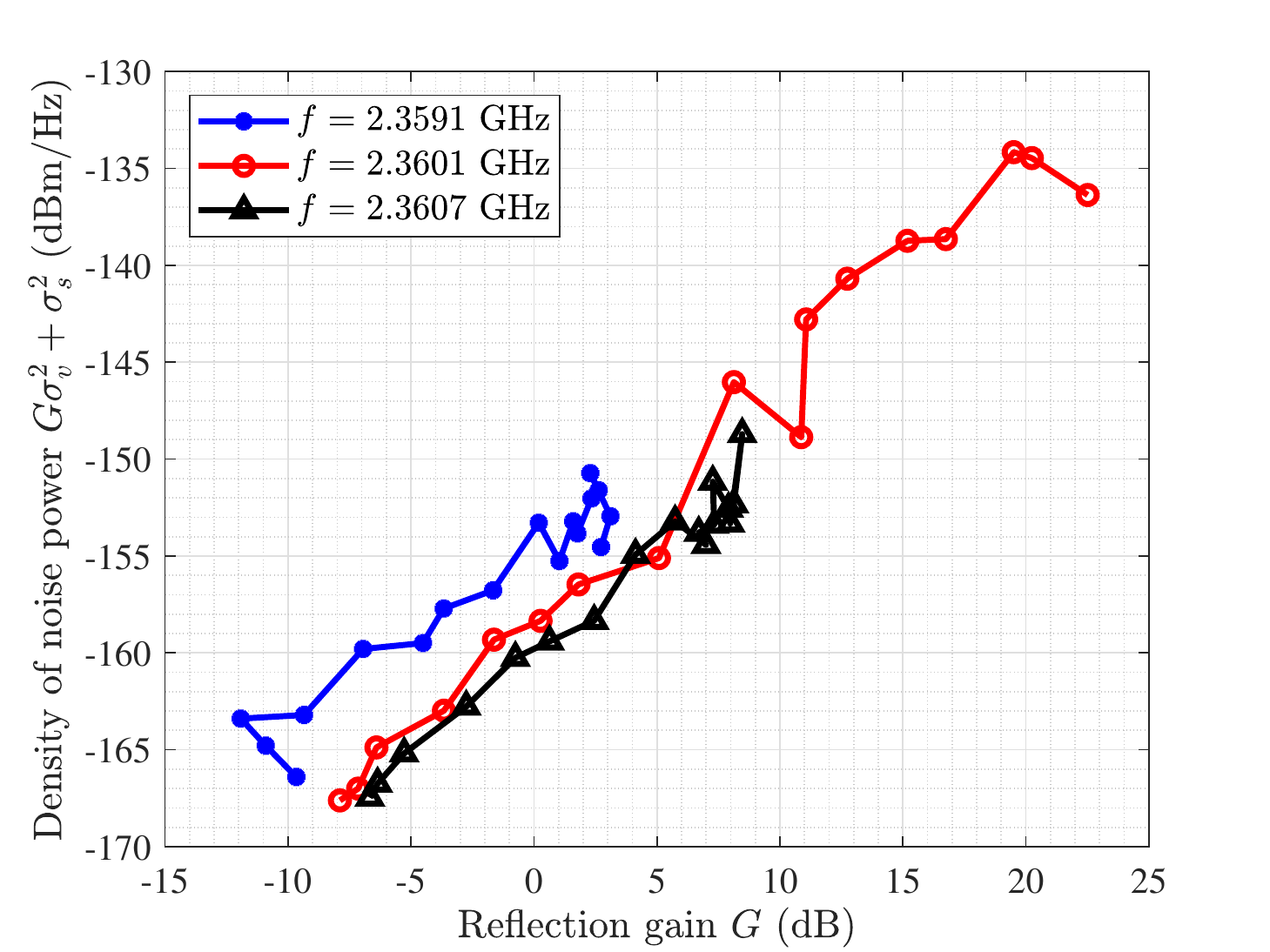}
	\vspace*{-1em}
	\caption{Experimental measurement result for the density of noise power $G\sigma_v^2+\sigma_s^2$ versus reflection gain $G$.}		
	\label{img:val:signal2}
	\vspace*{-1em}
\end{figure}
\subsubsection{Hardware platform}
To validate the model in (\ref{eqn:active_model_power}), we first establish the hardware platform used for our experimental measurements in Fig. \ref{img:val:environment}. Due to space constraints, we refer the reader to the journal version of this paper \cite[Fig. 4]{Zhang'21} for detailed information about the hardware platform.

\subsubsection{Reflection gain measurement}
Using the measurement system for the reflection gain depicted in \cite[Fig. 4 (b)]{Zhang'21}, we first investigate the reflection gain $G$ of the active RIS element. Note that the reflection gain $G$ can be reconfigured by the input power of the pump source $P_{\rm p}$. By setting the input power of the vector network analyzer as $P_x=-50$ dBm, the reflection gain $G$ as a function of the signal frequency can be directly measured via a vector network analyzer. Then, in Fig. \ref{img:val:signal1}, we show the measurement results for reflection gain $G$ as a function of signal frequency $f$ for different input powers of the pump source $P_{\rm p}$. We observe that the active RIS element can achieve a reflection gain $G$ of more than 25 dB, when $P_{\rm p}=18.24$ dBm, which confirms the significant reflection gains enabled by active RISs. 
\subsubsection{Noise power measurement}
We further study the noise power introduced by the active RIS element, i.e., ${G}\sigma _v^2 + \sigma _s^2$ in (\ref{eqn:active_model_power}), where ${G}\sigma _v^2$ and $\sigma _s^2$ are the powers of the dynamic noise and the static noise introduced at the active RIS element, respectively. Using the noise measurement system in \cite[Fig. 4 (c)]{Zhang'21}, we show the measurement results for the spectral density of noise power $G\sigma_v^2+\sigma_s^2$ as a function of $G$ for different operating frequencies in Fig. \ref{img:val:signal2}. We can observe that the noise power increases nearly linearly with $G$, which verifies the noise model ${G}\sigma _v^2 + \sigma _s^2$ in (\ref{eqn:active_model_power}). Particularly, for $f=2.3601$ GHz, the spectral density of $\sigma_s^2$ is about $-174$ dBm/Hz, while that of $\sigma_v^2$ is about $-160$ dBm/Hz, which is about $15$ dB higher. The reason for this is that the input noise is amplified by the noise factor, and additional noises are also introduced by the other active components such as the DC source used to power the active RIS.

\subsection{Simulation Results for Sum-Rate}\label{sub:vr:2}
\par	
\subsubsection{Simulation setup}
\begin{figure}[!t]
	\centering
	\includegraphics[width=0.45\textwidth]{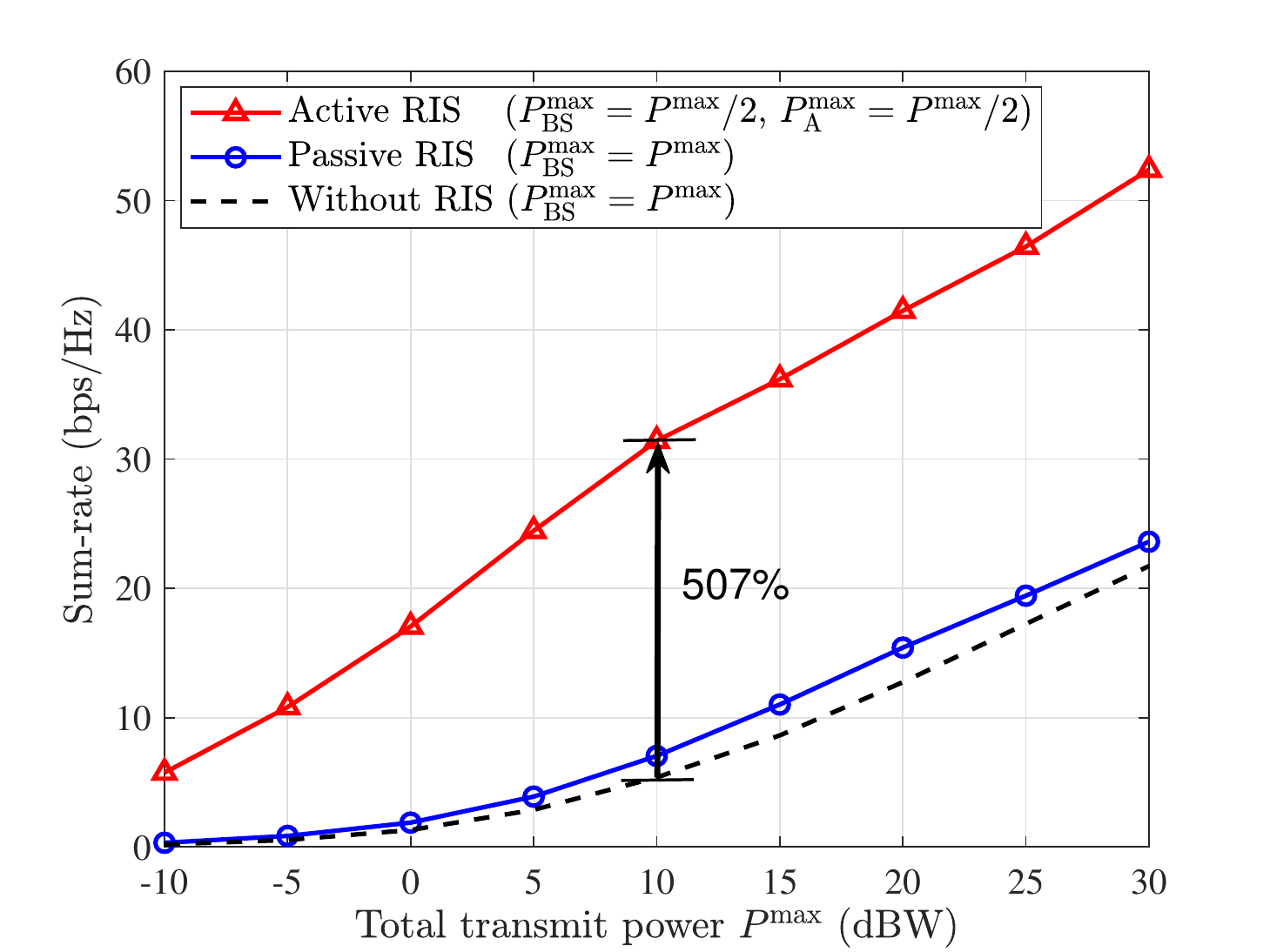}
	\vspace{-0.5em}
	\caption{Simulation results for the sum-rate versus total power consumption $P^{\max}$ in scenario 1 with a weak direct link..}	
	\label{img:val:sumratevsP1}
	\vspace{-1em}
\end{figure}	
We consider an active RIS aided MU-MISO system. Particularly, we consider two scenarios with different channel conditions. In scenario 1, the direct link is weak due to severe obstruction, while the direct link is strong in scenario 2. To be specific, two different path loss models from the 3GPP TS 36.814 standard are utilized to characterize the large-scale fading of the channels: i) ${\rm{P}}{{\rm{L}}_{s}} = 37.3 + 22.0\log d$; ii) ${\rm{P}}{{\rm{L}}_w} = 41.2 + 28.7\log d$, where $d$ is the distance between two devices. Path loss model ${\rm{P}}{{\rm{L}}_w}$ is used to generate the weak BS-user link in scenario 1, while ${\rm{P}}{{\rm{L}}_s}$ is used to generate the strong BS-user link in scenario 2. For both scenarios, ${\rm{P}}{{\rm{L}}_s}$ is used to generate the BS-RIS and the RIS-user channels. To account for small-scale
fading, we adopt the Ricean fading channel model for all
channels involved and we assume the Ricean factor as $\kappa=1$. 

The \ac{bs} and the active/passive RIS are located at (0, -60 m) and (200 m, 30 m), respectively. Four users are randomly located in a circle with a radius of 5 m from the center (200 m, 0). The numbers of BS antennas and RIS elements are set as $M=4$ and $N=256$, respectively. The noise power is set as ${\sigma}^2={\sigma}_v^2=-70$ dBm. For fair comparison, we constrain the total power consumption $P^{\max}:=P_{\text{BS}}^{\max}+P_{\text A}^{\max}$. For the active RIS, {\bf Algorithm 1} is employed for joint precoding and beamforming design, while for the passive RIS, the algorithm from \cite{Pan'19} is adopted.

\subsubsection{Simulation results}

\begin{figure}[!t]
	\centering
	\includegraphics[width=0.45\textwidth]{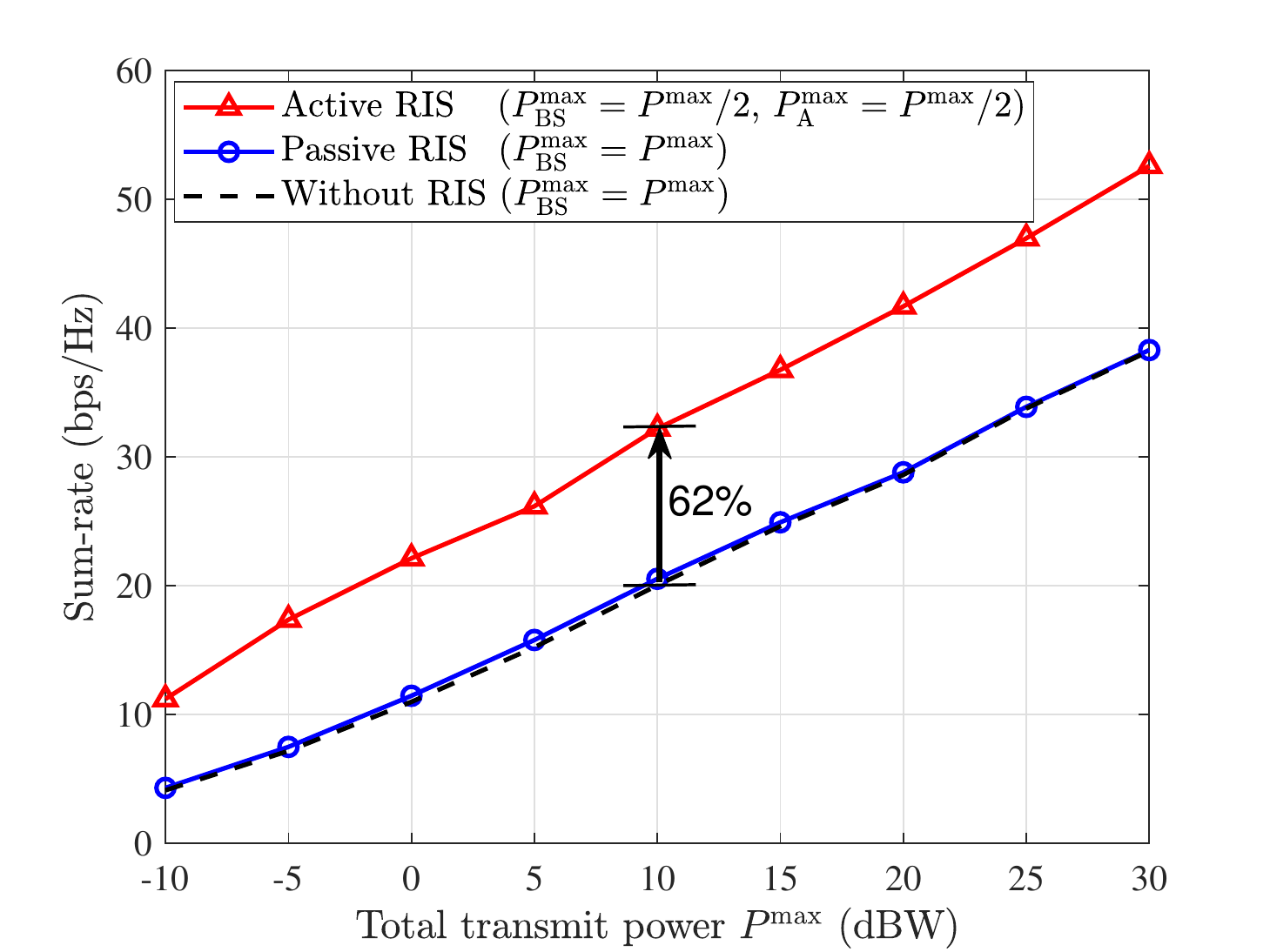}
	\vspace{-0.5em}
	\caption{Simulation results for the sum-rate versus total power consumption $P^{\max}$ in scenario 2 with a strong direct link.}	
	\label{img:val:sumratevsP2}
	\vspace{-1em}
\end{figure}

In Fig. \ref{img:val:sumratevsP1} and Fig. \ref{img:val:sumratevsP2}, we plot the sum-rate versus the total consumed power $P^{\max}$ for the two considered scenarios, where the direct link is weak and strong, respectively. Firstly, in scenario 1 with a weak direct link, the passive RIS can indeed achieve a performance improvement, while the active RIS achieves a much higher sum-rate gain. Secondly, in scenario 2 with a strong direct link, the passive RIS achieves only a negligible sum-rate gain, while the active RIS still realizes a noticeable sum-rate gain. For example, when $P^{\max}=10$ dBW, the capacities without RIS, with passive RIS, and with active RIS in scenario 1 are 5.34 bps/Hz, 7.00 bps/Hz, and 32.41 bps/Hz respectively, while in scenario 2, these values are 19.87 bps/Hz, 20.51 bps/Hz, and 32.18 bps/Hz, respectively. In this case, the passive RIS provides a 31\% gain in scenario 1 and a negligible 3\% gain in scenario 2. By contrast, the active RIS achieves noticeable sum-rate gains of 507\% in scenario 1 and 62\% in scenario 2, which are much higher than those achieved by the passive RIS.

\subsection{Field Test for a 64-Element Active RIS Aided Wireless Communication Prototype}
\begin{figure}[!t]
	\centering
	\includegraphics[width=0.45\textwidth]{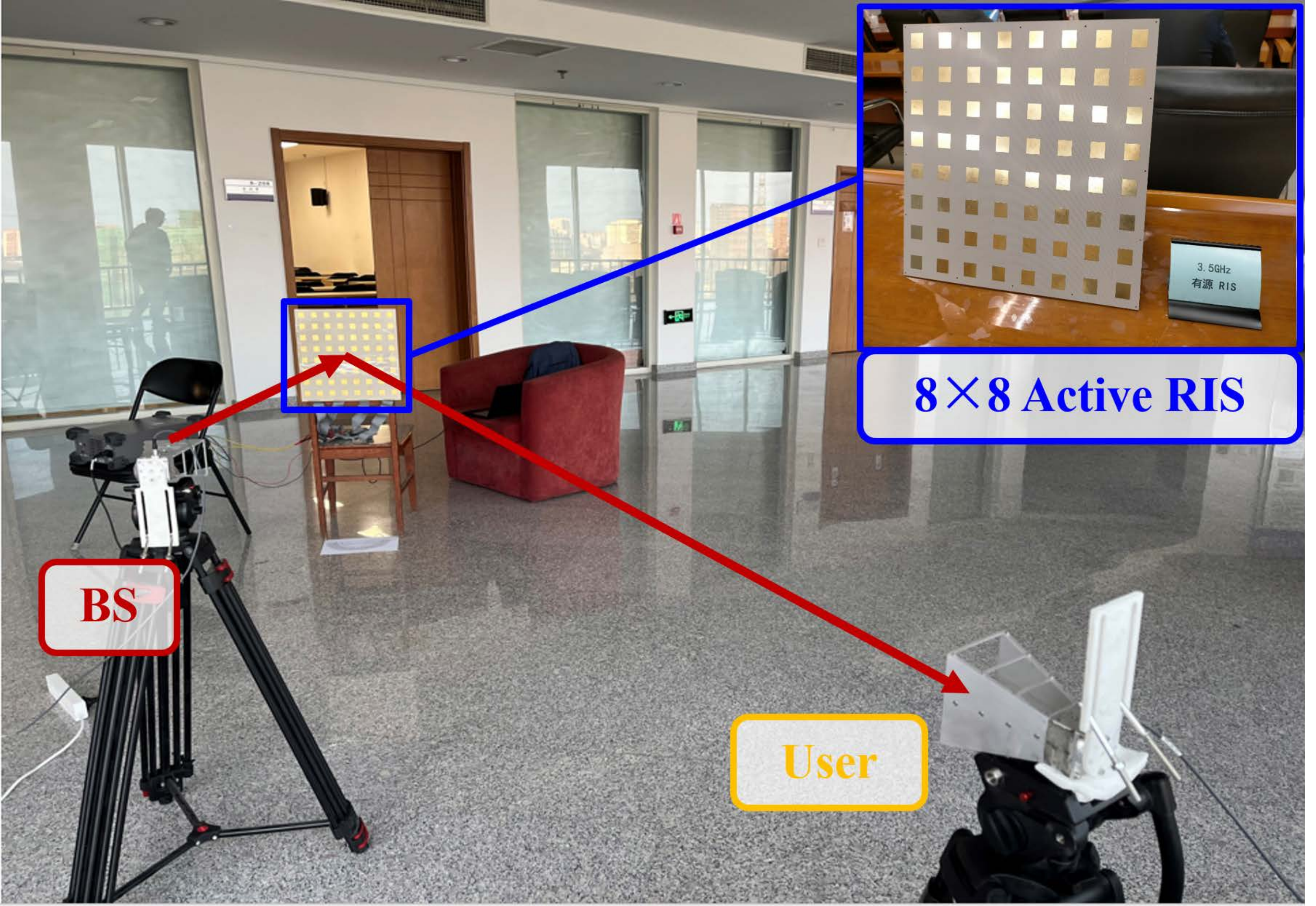}
	\vspace{-0.5em}
	\caption{A photograph of the developed 64-element active RIS aided wireless communication system.}	
	\label{img:val:Prototype}
	\vspace{-1em}
\end{figure}	
\subsubsection{64-element active RIS aided communication prototype}
To validate the significant gain of active RISs, we develop a 64-element active RIS aided wireless communication prototype, as shown in Fig. \ref{img:val:Prototype}. Specifically, the hardware structure of this prototype consists of three parts including a BS, a 64-element active RIS, and a user. For the BS and the user, two horn antennas with 13 dBi antenna gain are used to transmit and receive the signals, and the universal software radio peripherals (USRPs) are deployed to generate and process the baseband and RF signals (hardware version: USRP-2953R). By periodically expanding the active RIS elements designed in \cite{Xibi'20}, the 64-element active RIS is an 8$\times$8 plane array, of which each element has a reflection gain of $G=10$ dB.

\subsubsection{Experimental environment}
Based on the developed prototype, we establish the experimental environment for further validation. To match the transceivers, we configure the operating frequency of the active RIS to $f=3.5$ GHz and the bandwidth to 40 MHz by adjusting the circuit impedance of active elements. The polarization of the antenna at the BS and that at the user are selected as vertical and horizontal, respectively. The transmit power is set to $-10$ mW. We fix the heights of the BS, the RIS, and the user as 1 m. The horizontal distance of the BS-RIS link and that of the RIS-user link are set to 2 m and 3.5 m, respectively. The angle of arrival (AoA) at the active RIS is fixed as $0^{\circ}$, and the angle of departure (AoD) will be specified to evaluate the performance gain of active RISs at different orientations. To observe the reflection gain of the active RIS, we use a metal plate with the same aperture size as the active RIS for performance comparison.

\subsubsection{Experimental results}
By moving the user at different AoDs and configuring the phase shift of the active RIS with discrete Fourier transform (DFT) codebook, we obtain the experimental results shown in Table I. One can observe that, compared with the received power for the metal plate, the active RIS can always achieve a gain of about $10$ dB. The data rate for the active RIS can hold at about 30 Mbps, while that for the metal plate only ranges from 1 Mbps to 2Mbps. The reason is that, the beamforming at the active RIS can make the reflected beam with high array gain and reflection gain, while the metal plate can only reflect the incident signals randomly without in-phase combination or amplification, which validates the significant gain of active RISs.

\begin{table}
	\centering
	\small
	\caption{Experimental results for the developed prototype}
	\vspace*{-0.5em}
	\begin{tabular}{|c|c|c|c|} 
		\hline
		\hline
		\textbf{AoD}         & \textbf{Device} & \multicolumn{1}{c|}{\textbf{Received Power}} & \multicolumn{1}{c|}{\textbf{Data Rate}}  \\ 
		\hline
		\multirow{2}{*}{$15^{\circ}$} & Metal plate     & -110 dBm                                     & 1.2 Mbps                                  \\ 
		\cline{2-4}
		& Active RIS      & -100 dBm                                     & 28.5 Mbps                                 \\ 
		\hline
		\multirow{2}{*}{$30^{\circ}$} & Metal plate     & -105 dBm                                     & 1.5
		Mbps                                 \\ 
		\cline{2-4}
		& Active RIS      & -98 dBm                                      & 30.5
		Mbps                                \\ 
		\hline
		\multirow{2}{*}{$45^{\circ}$} & Metal plate     & -105 dBm                                     & 1.5
		Mbps                                 \\ 
		\cline{2-4}
		& Active RIS      & -95 dBm                                      & 30
		Mbps                                  \\ 
		\hline
		\multirow{2}{*}{$60^{\circ}$} & Metal plate     & -108 dBm                                     & 2
		Mbps                                   \\ 
		\cline{2-4}
		& Active RIS      & -90 dBm                                      & 32
		Mbps                                  \\
		\hline
		\hline
	\end{tabular}
\vspace*{-1em}
\end{table}

\section{Conclusions}\label{sec:con}
In this paper, we have studied the concept of active RISs to overcome the fundamental limitation of the “multiplicative fading” effect. Firstly, we have verified the signal model of active RISs through the experimental measurements on a fabricated active RIS element. Based on the verified signal model, we have formulated the sum-rate maximization problem for an active RIS aided MU-MISO system and a joint precoding and beamforming algorithm has been proposed to solve this problem. Simulation results have shown that, in a typical application scenario, the existing passive RIS can realize only a negligible sum-rate gain of about 3\%, while the active RIS can achieve a substantial sum-rate gain of about 62\%, thus indeed overcoming the “multiplicative fading” effect. Finally, we have developed a communication wireless communication prototype aided by a 64-element active RIS, and the significant gain of active RISs is validated by field test. In the future, many research directions for
active RISs are worth pursuing, such as hardware design, prototype development, channel estimation, and energy efficiency analysis.

\section*{Acknowledgment}
This work was supported in part by the National Key Research and Development Program of China (Grant No. 2020YFB1805005), in part by the National Natural Science Foundation of China (Grant No. 62031019), and in part by the European Commission through the H2020-MSCA-ITN META WIRELESS Research Project under Grant 956256. 

\footnotesize
\bibliographystyle{IEEEtran}


\bibliography{IEEEabrv,reference}

\end{document}